\documentclass[fleqn,usenatbib]{mnras}
\usepackage{graphicx}
\usepackage{url}
\usepackage{longtable}
\usepackage{float}
\usepackage{subfigure}
\usepackage{newtxtext,newtxmath}
\usepackage[T1]{fontenc}
\usepackage{amsmath}
\usepackage{soul}
\usepackage{lscape}

\DeclareRobustCommand{\VAN}[3]{#2} \let\VANthebibliography\thebibliography
\def\thebibliography{\DeclareRobustCommand{\VAN}[3]{##3} \VANthebibliography}

\title[Kinematics of the molecular interstellar medium]{Kinematics of the molecular interstellar medium probed by Gaia: steep velocity dispersion-size relation,  isotropic turbulence, and location-dependent energy dissipation}

\author[Zhou, Li \& Chen]{
Ji-Xuan Zhou,$^{1}$ 
Guang-Xing Li,$^{1}$\thanks{gxli@ynu.edu.cn} 
Bing-Qiu Chen,$^{1}$\thanks{bchen@ynu.edu.cn} \\
$^{1}$ South-Western Institute for Astronomy Research, Yunnan University, Chenggong District, Kunming 650091, P.\,R. China}

\begin{document}
\label{firstpage}

\maketitle

\begin{abstract}
    The evolution of the molecular interstellar medium is controlled
    by processes such as turbulence, gravity, stellar feedback, and Galactic shear. AL a part of the \href{https://gxli.github.io/ISM-6D/}{ISM-6D} project, using Gaia astrometric measurements towards a sample of young stellar objects (YSOs),  we study the morphology and kinematic structure
    of the associated molecular gas. We identify 150 YSO associations 
    with distance $d \lesssim 3 \;\rm kpc$.
     The YSO associations are elongated, with a median aspect ratio of 1.97, and are
    oriented parallel to the disk midplane, with a median angle of 30$^{\circ}$. The turbulence in the molecular clouds as probed by the YSOs is isotropic, and the velocity dispersions are related to the sizes by $\sigma_{v,{\rm 2D}}  = 0.74\;(r/{\rm pc})^{0.67} \;({\rm km/s})\;$.
    The slope is on the steeper side, yet consistent with previous
    measurements. The energy dissipation rate of turbulence $\dot{\epsilon} =
    \sigma_{v,{\rm 3D}}^3 /L$ decreases with the Galactocentric distance, with a gradient of 0.2 $\rm dex
    \; kpc^{-1}$, which can be explained if turbulence is driven by cloud collisions. In this scenario,  the 
    clouds located in the inner Galaxy have higher chances to accrete smaller clouds and are more turbulent. 
     Although the density structures of the complexes are
    anisotropic, the turbulence is consistent with being isotropic. 
     If the alignment between density structures and the Galactic-disk mid-plane is due to shear, we expect $t_{\rm cloud} \gtrsim t_{\rm shear}\approx 30\; \rm Myr$. This cloud lifetime is longer than the turbulence crossing time, and a continuous energy injection is required to maintain the turbulence. 
\end{abstract} 

\begin{keywords} 
    Galaxies: ISM  -- ISM: structure -- ISM: kinematics and dynamics  -- Stars: formation  --  Physical data and processes: turbulence
\end{keywords}

\section{Introduction}
The collapse of molecular clouds is a complex processes characterized by 
interplays between e.g. gravity, turbulence \citep{1981Turbulence} and magnetic
field \citep{2014prpl.conf..101L}. The clouds exhibit complex density and
velocity structures. \cite{1981Turbulence} found the velocity dispersion
$\sigma_{v}$ of a cloud is positively correlated with its size $l$ by $\sigma_{v}
\sim l^{\beta}$ where $\beta = 0.38$. The slope can be explained by the cascade
of Kolmogorov-like turbulence \citep{1941Kolmogorov}. This relation is called
the Larson relation, and the role of turbulence in star formation is
well-recognized ever since. However, the slope of the Larson relation remains
uncertain, and the interpretation of turbulence cascade has been challenged.
\citet{1987Solomon} found that the velocity dispersion $\sigma_{v}$ of a cloud
is related to its size $l$ by $\sigma_{v} \sim l^{0.5}$ from 273 molecular
clouds observed by the Five College Radio Astronomy Observatory (FCRAO), where
the scaling exponent is larger than 0.38 as found by \cite{1981Turbulence}. Similar 
results are also reported in e.g. \citet{2008Bolatto}.
\citet{2009Heyer,2011Ballesteros} proposed that the slope of the Larson relation can also be explained 
assuming self-gravity. Recently, \cite{2021Izquierdo} simulated the evolution of molecular clouds in
 galaxy discs, and found that the exponent varies from 0.3 to 1.2. 


The fact that larger velocity dispersions are found at larger scales suggests
that the kinetic energy of molecular clouds is injected from the outside. To
constrain turbulence injection, a promising direction is to identify coherent
gas structures. \citet{2013A} identified a kpc-sized coherent structure called
the “500 pc filamentary gas wisp”,  and \cite{2014Goodman} discovered an object
called “extended Nessie” of a similar size. These discoveries suggest that the
molecular clouds are not separated objects but an inseparable part of the Milky
Way interstellar medium. Subsequent works studied samples of such filaments
\citep{2014A&A...568A..73R, 2016Ragan, 2015Wang,2016li}. \cite{2016li} extracted
filamentary structures from the APEX Telescope Large Area Survey of the Galaxy
\citep[ATLASGAL,][]{2009A&A...504..415S} survey. They found that the filamentary
structures  tend to stay parallel to the Galactic disk mid-plane, indicating
that Galactic shear plays an important role in shaping the structures of the
molecular interstellar medium. Another possible consequence of shear is that it
can lead to velocity anisotropy, e.g. the velocity dispersion of molecular
clouds measured along the Galactic longitude $l$ direction is larger compared to that along the Galactic latitude $b$ direction.
 \cite{2006jin} attempted to constrain the importance of
shear by studying the radial velocity structures of the Boston University–Five
College Radio Astronomy Observatory Galactic Ring Survey (GRS) clouds
\citep{2006ApJS..163..145J}, and found that the velocity gradient has no
preferred direction. Whether the velocity field is isotropic or anisotropic
remains unclear. 

We study the properties and Galactic distribution of a sample of YSO
associations using Gaia data. YSOs are stars at the very early stage of stellar
evolution and are usually close to where they were born, such that they inhibit
the velocity of gas from which they originate and their locations may indicate
the shape of molecular clouds. \cite{ORION_COMPLEX,2021PASJ...73S.239L} compared
the velocities of YSOs against the velocity of the ambient molecular gas, and
found a good agreement. This technique has already been used in previous studies
to study e.g. turbulent motion of molecular gas \citep{2021ha}. Li \& Chen (2021
under review) used a similar approach to study the velocity structure of a
Galactic filament called the Gould Belt Radcliffe Wave
\citep{2020Natur.578..237A}. In this paper, most YSO associations have sizes
similar to that of the clouds, in most cases, they are associated with
molecular clouds (Zhou, Li et al in prep). The Gaia survey
\citep{2016A&A...595A...1G} provides accurate measurements of the magnitudes,
parallaxes, and proper motions of stars in the Milky Way. We use the Gaia Data Release 2
\citep[DR2,][]{GAIADR2} measurements of young stellar objects to study the
kinematics of the gas they associate with. Using a sample of 15149 YSOs which
are younger than 3 Myr and located from $-$6 to $-$10 kpc from the Galactic
center, we identify a sample of $\sim$100 molecular cloud-sized YSO associations and study the kinematics of the molecular interstellar medium from several pc to one
hundred pc scale. 


\section{Data \& Method}
\subsection{YSO data}
Our YSO sample, which consists of 15149 Class\
\uppercase\expandafter{\romannumeral1} and Class\
\uppercase\expandafter{\romannumeral2} YSOs, is selected from \citet{2016An}. \citet{2016An}
obtained those YSOs by appliying  the Support Vector Machine method to a combined dataset
containing the AllWISE catalog of the Wide-field Infrared Survey Explorer
\citep{2014allwise}, the Two Micron All-Sky Survey  \citep[2MASS][]{20032mass},
and the Planck dust opacity values \citep{2014planck}. Their parallexes and
proper motions are derived by crossing-matching with Gaia DR2
\citep{GAIADR2}. According to \citet{YSOage}, age for Class\
\uppercase\expandafter{\romannumeral1} YSO is about 0.78 Myr and age for Class\
\uppercase\expandafter{\romannumeral2} YSO is 2 or 3 Myr. As the timescale is shorter than the crossing time of molecular cloud turbulence,
we expect them to inherit the positions and velocities of 
the gas they originate from  (Li \& Chen 2021 under review). We further remove  YSOs  whose
parallax errors are larger than 20\%.

\subsection{Estimating the spatial density}
\begin{figure*}
\begin{center}
\includegraphics[scale=1.1]{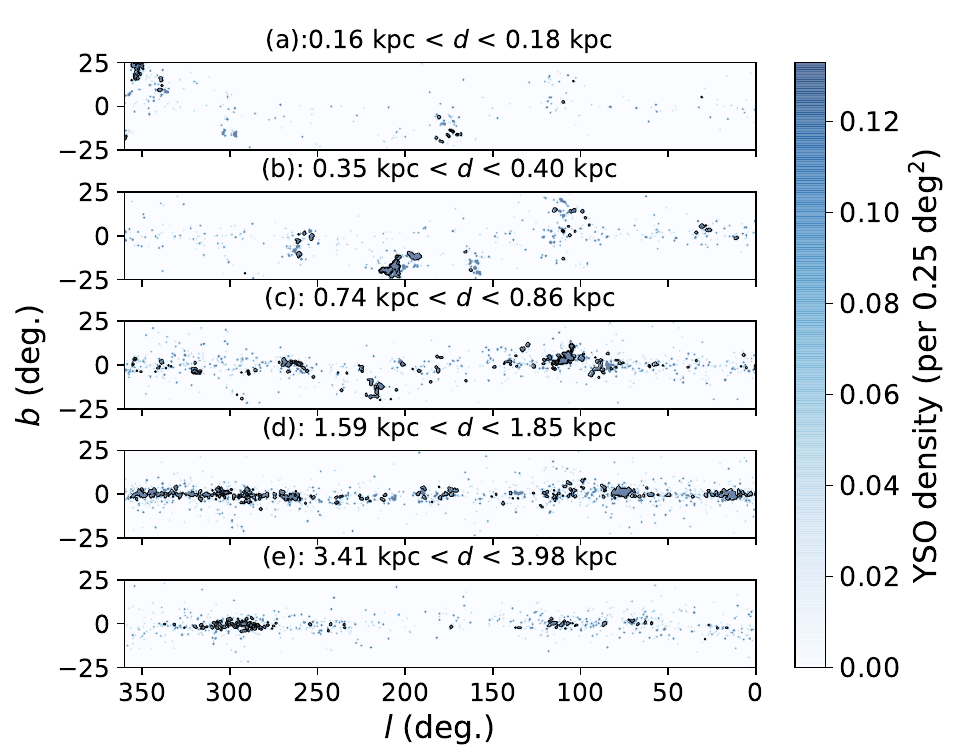}  
\caption{{\bf Spatial density of the YSO sample at five distance channels.} The colors represent the YSO density, and the black contours represent the boundaries of the associations found by \texttt{Dendrogram}.}
\label{fig1}
\end{center}
\end{figure*}

We use the \texttt{python} function \texttt{hist3d} to produce a YSO density map
in $l$, $b$, and $\rm log$($d$) space, where
$d$: the distance to the Sun. Our density map has  $ 0^{\circ} < l < 360
^{\circ}$,  $ -25^{\circ}< b <  25 ^{\circ}$,  3.5 pc  $<d<$ 7300
pc. We divide $l$ into 720 bins separated by 0.5$^{\circ}$, $b$ into 100 bins
separated by 0.5$^{\circ}$ and $\rm log$($d$) into 50 bins. The density map is
smoothed with a Gaussian filter with sigma = 0.6 pixel. The results are plotted
in Fig.~\ref{fig1}.

\subsection{Identifying structures}
We use the \texttt{Dendrogram} program \citep{Rosolowsky2008} to extract
structures from the YSO density map. \texttt{Dendrogram} is a program to extract
structures from  maps by exploiting the topological relation between
isosurfaces. The results are dependent on three parameters: \texttt{min\_value},
\texttt{min\_delta}, and \texttt{min\_npix}, where \texttt{min\_value} means the
minimum value to be considered as a structure, \texttt{min\_delta} refers to the
value difference between structures and substructures, and \texttt{min\_npix} is
the smallest number of pixels contained in one structure.

\begin{figure*}
\begin{center}
\includegraphics[scale=0.77]{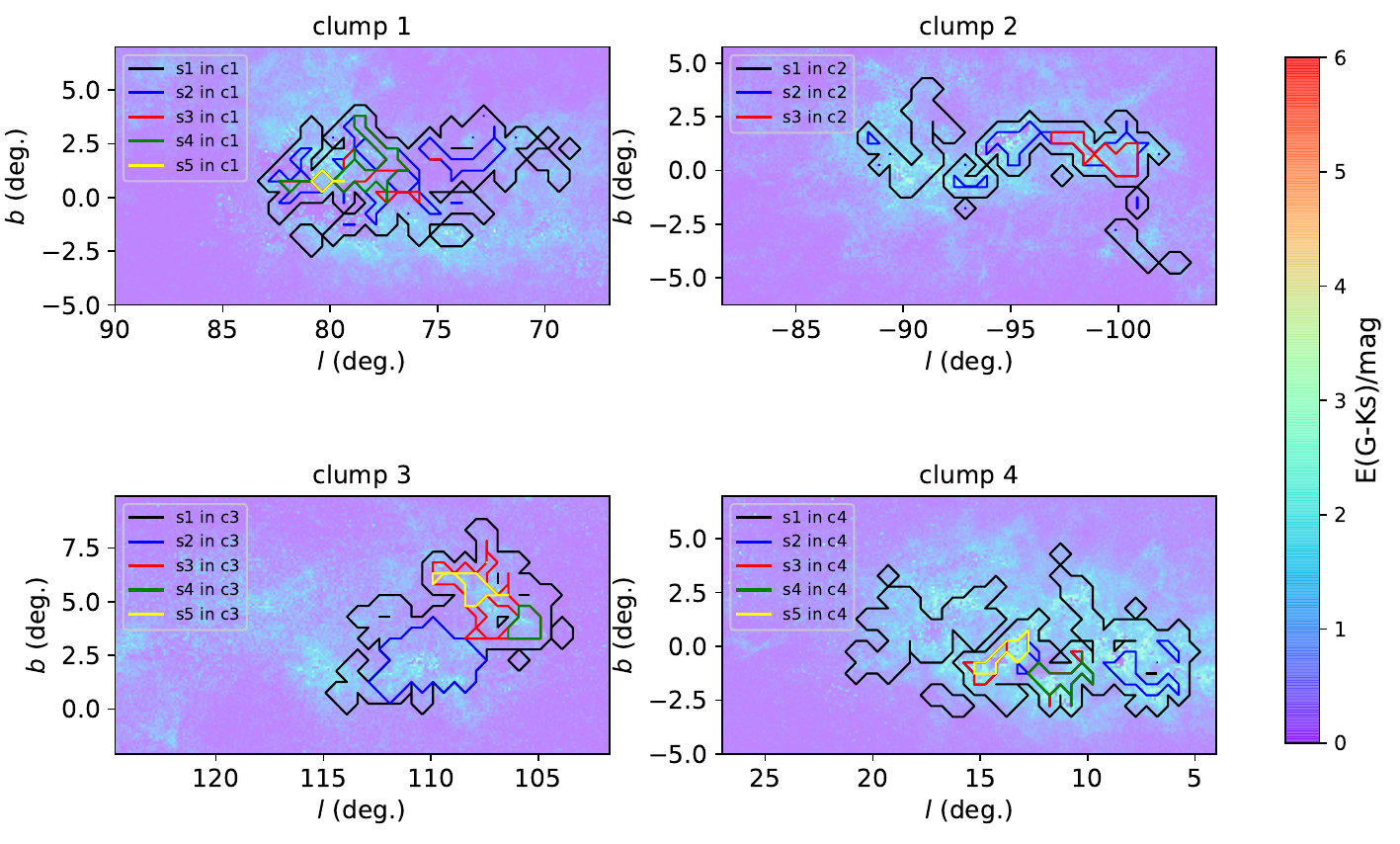}
\caption{{\bf Examples of nested structures.} The background is the 2D dust map from \citet{2019MNRAS.483.4277C} integrated from $d = d_{\rm mean } - \delta_d$ to $d = d_{\rm mean } + \delta_d$, where $d_{\rm mean}$ is the mean distance, and $\delta_d$ is the widths. Contours represent boundaries of the associations.}
\label{fig2}
\end{center}
\end{figure*}

Through experiments, we set \texttt{min\_value} $= 0.1$, \texttt{min\_delta} $=
0.3$, and \texttt{min\_npix} $= 10$, and obtain a total of 518 structures, which
contain 10 to 2593 pixels. One pixel might belong to a few nested structures.
All structures found by \texttt{Dendrogram} are called ``associations''. Some of
the small associations are part of the larger ones, and a few examples are shown
in Fig.~\ref{fig2}. Each YSO can belong to one or a few associations, and their
memberships are determined by matching their locations to the footprints of
identified structures. We further remove associations
that contain very few stars($\le 15$), and obtain a sample of  219
associations which are coherent in space and contain  reasonable numbers of stars.

\begin{figure}
\begin{center}
\includegraphics[scale=0.165]{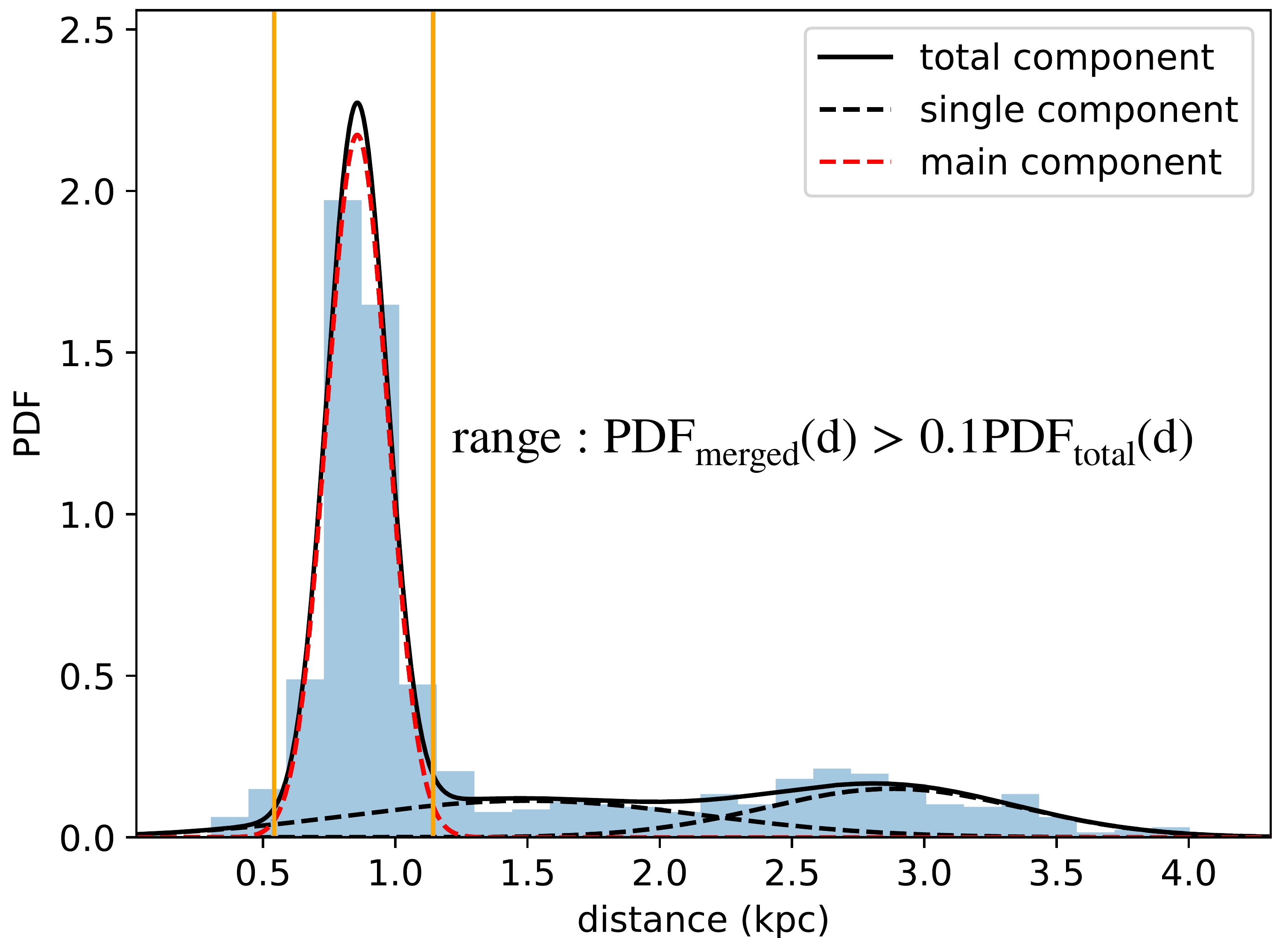}   
\caption{{\bf Generate a clean sample of member YSOs.} The histogram represents the distance distribution of YSOs that belongs to one of our complexes. The solid black line refers to the overall best-fitting distribution selected from a range of models based on the AIC, and the dashed lines represent different Gaussian components. These are derived using \texttt{sklearn.mixture}. The red dashed line is the  dominant merged component, and the two vertical orange lines mark the final distance range. YSOs that stay within this distance range will be considered in our analyses.}
\label{fig3}
\end{center}
\end{figure}

\subsection{Identifying member YSOs}
\label{sec:clean} Almost all YSO
associations contain significant amounts of contaminants, as reflected from
their distance distributions. To address this, we  apply a Gaussian Mixture
Model (\texttt{sklearn.mixture}) to the distance distributions. The component
number is allowed to vary from 1 to 4, after which the best-fitting one is
selected based on the Akaike Information Criterion (AIC), which describes the
simplicity and precision of the model. This procedure leaves us with a number of
Gaussian components named $G({\rm d}_i, {\rm Ed}_i)$, where ${\rm d}_i$ is the
mean distance and ${\rm Ed}_i$ is the distance dispersion. We first select the
most dominant components as $G({\rm d}_{\rm p}, {\rm Ed}_{\rm p})$. If there is
an additional component staying close to the most dominant one ($|d_i - d_{\rm p
}| < 1.6\, {\rm Ed}_{\rm p}$ ), they will be merged together, after which a
valid distance range is determined as where $ {\rm PDF}_{\rm merged} (d)  > 0.1\;
{\rm PDF}_{\rm total} (d)$, as indicated in Fig. \ref{fig3}, where PDF stands for the probability density distributions. A YSO is a member of the association if its distance
is within the range as determined by this procedure. If a significant fraction
($\gtrsim $65\%) of the member YSOs do not belong to the merged component, the
association is removed from the sample. An example illustrating this procedure
is plotted in Fig. \ref{fig3}.

\subsection{Final sample}
After removing the contaminations in associations, we apply the following criteria to obtain a sample of YSO associations whose
physical properties can be estimated reliably: (1) the number of stars is more
than 20,  and (2) the dispersion of distance is smaller than 0.35 kpc. Finally, we obtain
 a sample of 150 YSO associations, which are described in Appendix A.

\begin{figure}
\begin{center}
\includegraphics[scale=0.52]{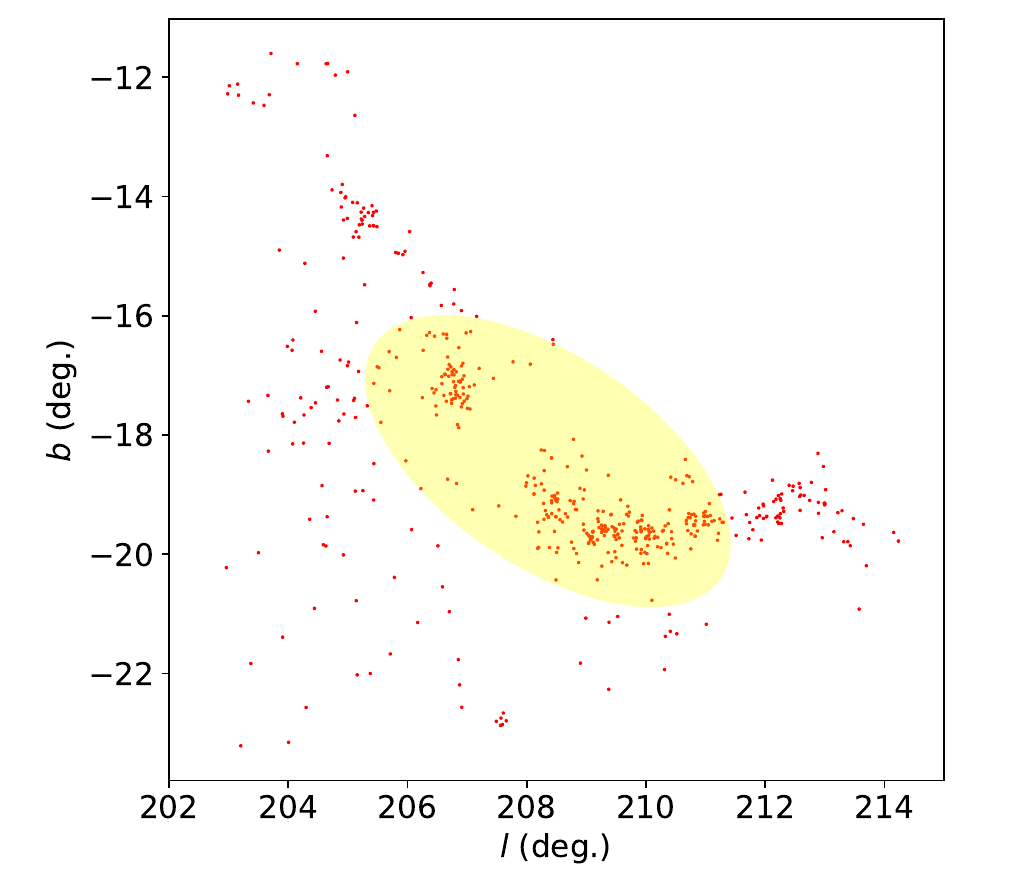}
\includegraphics[scale=0.51]{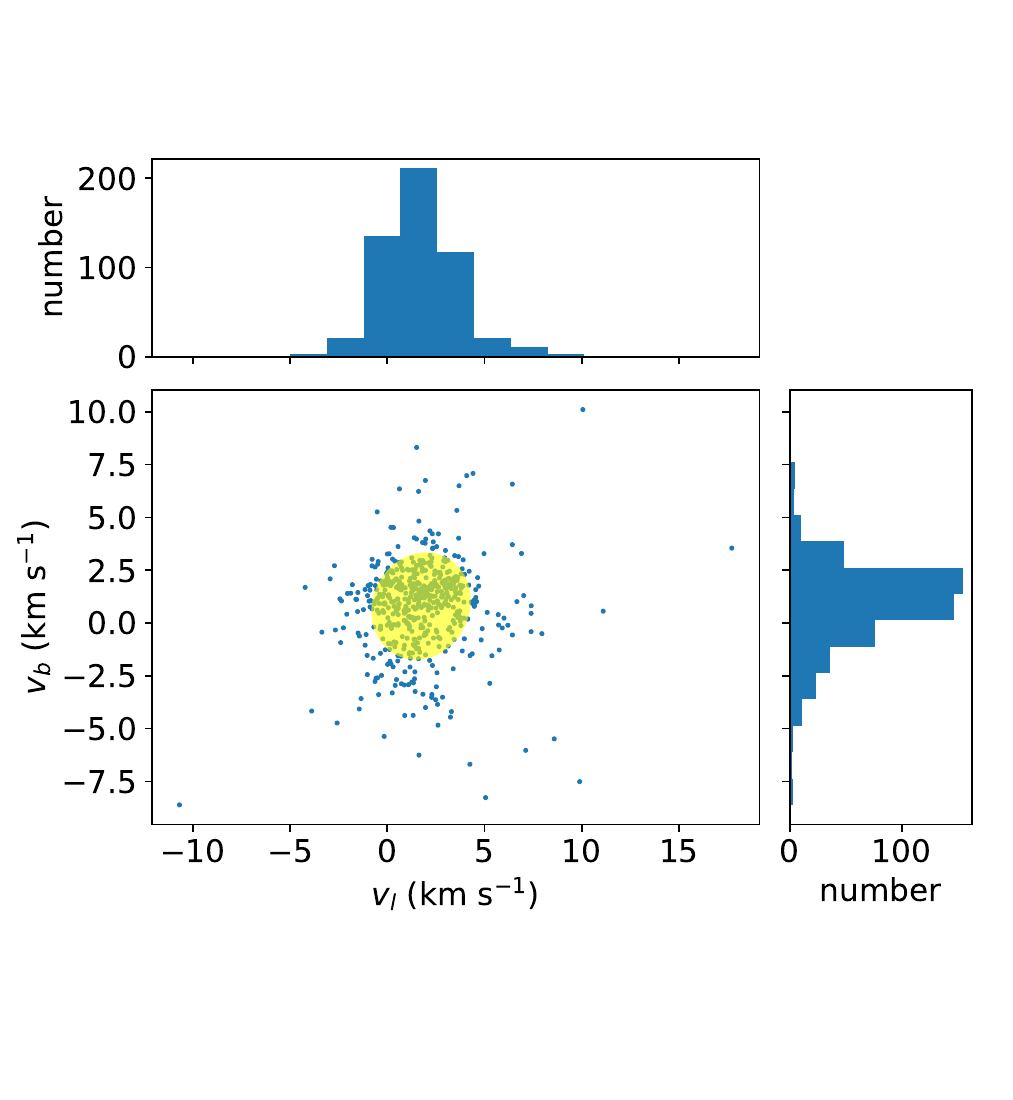}   
\caption{{\bf Locations of the member YSOs of one of our associations in the $l-b$ and $v_l - v_b$ space. Upper panel:} Locations  in the $l-b$ plane.
Red dots are the member stars of this structure and the yellow ellipse represents the fitting result, as described in section.\ref{sec:1}, through which the aspect ratio and the orientation are measured. {\bf Lower panel:}
Locations of member YSOs in the 
$v_l - v_b$ space. The yellow ellipse represents the fitting result, where the heights of the ellipses are $2 \; r_{\rm min}$.}
\label{fig4}
\end{center}
\end{figure}

The member stars have spatial distributions that resemble the molecular gas they
originate from, and in the majority of the cases, their velocity distributions
are Gaussian-like. One example is presented in Fig.~\ref{fig4}, which
corresponds to the well-known Orion molecular clouds. Here, we plot the
distributions of member stars in the $l-b$ and $v_l-v_b$ space. In the $l-b$
space, the YSO association resembles the Orion molecular cloud, and in the
velocity space, the YSOs are clustered.

\subsection{Deriving physical properties}
For each YSO association, we derive the following properties.

\subsubsection{Distance and velocity dispersion}
We use the Random Sample Analysis to estimate the distance and velocity uncertainties. 
For each proper motion and distance measurements, assuming the errors follow a Gaussian distribution, we generate a number of realizations according to the measured uncertainties in parallax and  proper motion, from which distances and proper motions as well as the statistical uncertainties can be derived. 


In practice, the measured parameters include the proper motions and
parallaxes of its member stars. We start by generating $N=1000$ new measurements
where we added uncertainties to the parameters according to the expected random
measurement errors. To measure the distance and its uncertainty, assuming the
mean distances of a realization is $d_{{\rm mean}, i}$, the mean distance is
${\rm mean} (d_{{\rm mean}, i})$, and the  uncertainty of distance estimate is
${\rm std}({\rm d}_ {\rm mean,i})$, where $\rm std$ stands for the standard deviation. Similar approaches can be used
to measure the velocity dispersion and its uncertainty. 

(Using the proper motion uncertainty and distance uncertainty, we generate 1000 new measurements for each stars' velocity in $l$ and $b$ directions and 2D velocity. Assuming the mean velocity of one velocity distribution in the 1000 realizations is $v_{{\rm mean},i}$ and the standard deviation of the velocity distribution is $\sigma_{v, i}$, the mean velocity for an association is mean($v_{{\rm mean},i}$), velocity dispersion $\sigma_v$ is mean($\sigma_{v, i}$), and velocity dispersion error E($\sigma_v$) is std($\sigma_{v, i}$).)

\subsubsection{Size}
The dataset allows us to study the motion along $l$ and $b$ directions
separately. To fully exploit this potential, we derive three sizes: size in $l$
direction - $r_{l}$, size in $b$ direction - $r_{b}$, and overall size $r$. These are full width at half maxima (FWHM) widths, e.g. $r_{l}$= 2.355 $d\cdot {\rm std} (l_i)$, and  $r_{b}$ = 2.355$d\cdot {\rm std} ({b_i})$, where $l_i$, $b_i$ are positions of the member YSOs in one association. The size $r$ is
estimated using: $r = 2.355d\cdot \sqrt{(\sigma_{\rm min}^2 + \sigma_{\rm max}^2)/ 2}$,  where
$r_{\rm min} = \sigma_{\rm min} \; d$ and $r_{\rm max} = \sigma_{\rm max} \; d$ , and $\sigma_{\rm min}$, $\sigma_{\rm max}$ are the dispersions measured along the major and minor axes of the spatial distributions on sky, and are the parameters we derive from the fitting ellipses in Sec.\ref{sec:1}. $d$
is the mean distance of every association. Our associations have sizes ranging from several pc to hundreds of pc. The intrinsic width of the cloud is estimated as width = $\sqrt{{\rm std}(d_{i})^{2} - {\rm std} (d\_{\rm err}_{i})^{2}}$ where $d_i$ is the distance of object $i$ and the corresponding error is $d\_{\rm err}_{i}$.

\begin{figure*}
\begin{center}
\includegraphics[width=\textwidth]{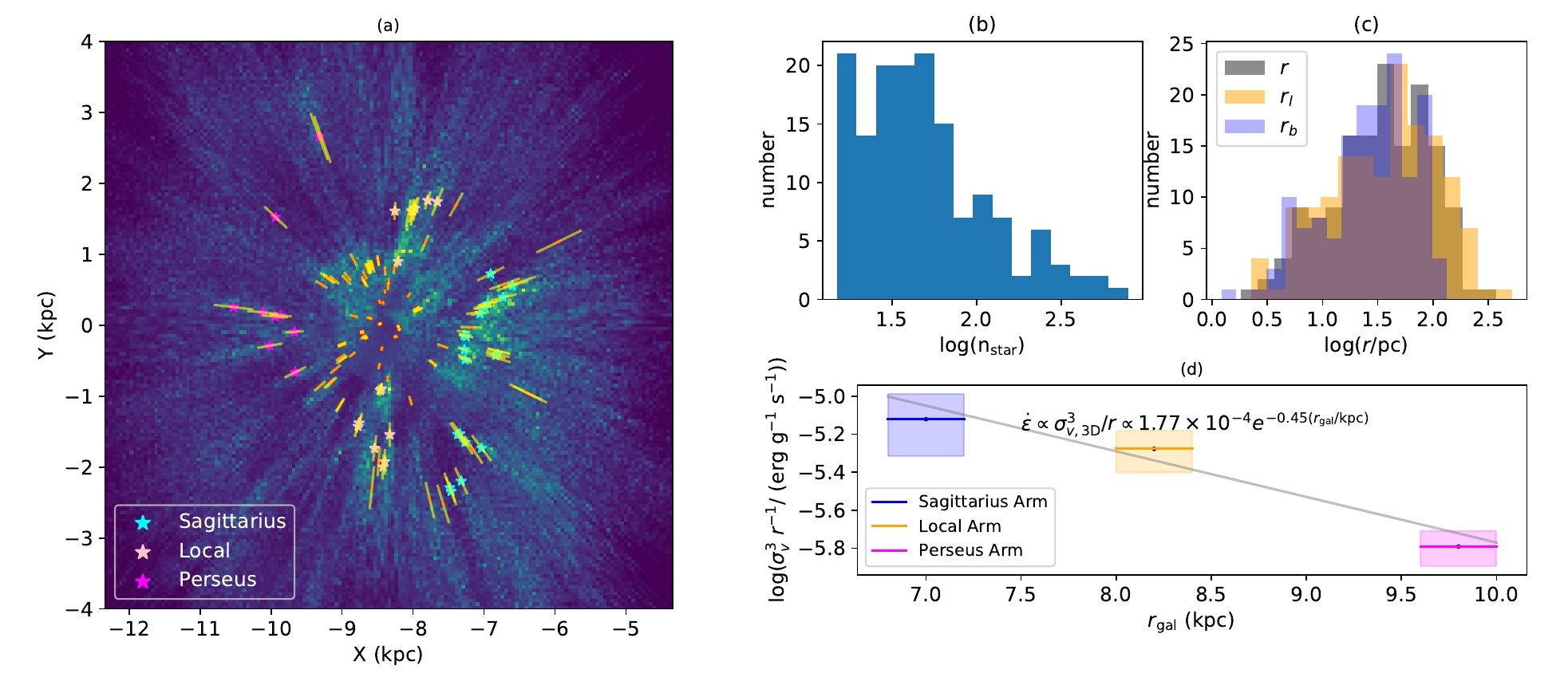} 
\caption{{\bf Properties, Galactic distribution, and energy dissipation rate of YSO associations. Panel (a):} Location of the associations on the X-Y plane. {The sun locates at (-8.34, 0) and the center of the Milky Way locates at (0, 0).}
Red dots
represent the locations of 150 associations and yellow lines show their distance uncertainties. The background dust map
is from \citet{2019MNRAS.483.4277C}. The cyan, pink, and fuchsia stars refer
to the associations located on the Sagittarius Arm,
the Local Arm, and the Perseus Arm respectively, and the bars represent distance uncertainties. {\bf Panel (b):} Histogram of
the number of member stars for all 150 associations. {\bf Panel (c):} Histogram of the size
of associations. The Grey one represents the distribution of the overall size, the yellow one represents the sizes measured along the Galactic longitude ($l$) direction, and the purple one represents the sizes measured along the Galactic latitude ($b$) direction. {\bf
Panel (d):} Energy dissipation rate for three spiral Arms. Fuchsia, orange and blue lines with error bars present the energy dissipation rate of associations located on the Perseus Arm, the Local Arm, and the Sagittarius Arm respectively, where the corresponding errors are also indicated using the extents of the boxes. The grey line is the fitting result where $\dot{\epsilon} = 1.77\times10^{-4}e^{-0.45(r_{\rm gal}/{\rm kpc})}({\rm erg\ g^{-1}\ s^{-1}})$.} 

\label{fig5}
\end{center}
\end{figure*}

\subsubsection{Aspect ratios and Orientations}\label{sec:1}
For each association, we fit ellipses to the distribution of member YSOs in the
spatial ($l-b$) and velocity ($v_l-v_b$) space by diagnosing a
tensor constructed from the spatial and velocity dispersions (e.g. the tensor is
$T_{\rm i,  j } = \sum p_i, p_j$, where $p$ is the spatial/velocity coordinates,
and $i, j$ represent $l$, $b$ directions), from which the lengths of the major and minor axes $\sigma_{\rm max}$, $\sigma_{\rm min}$ as well as their orientations can be determined. These allow us to measure the orientation of the object from this major axes as well as the
aspect ratios. Spatially, we measure the position angle ${\rm
PA}_{l-b}$ and the aspect ratio ${\rm A}_{l-b}$ and in the velocity space,
we also measure the position angle ${\rm PA}_{v_l - v_b }$ and
the aspect ratio ${\rm A}_{v_l - v_b }$.

\section{Results} 
\subsection{Galactic distribution and size} 
In Fig.~\ref{fig5} we plot  locations of the associations in the Milky Way disk,
where a map of the distribution of dust from \citet{2019MNRAS.483.4277C} is
overlaid. The complexes are associated with dust and are organized along a few
filaments. This is expected, as the YSOs are young and are associated with the
molecular cloud which contains dust. In Fig. \ref{fig5}, we plot the distribution of the number
of member stars contained in the complexes, as well the size distribution. The
associations have sizes that range from 10 to 100 pc, which is similar to the
sizes of molecular clouds obtained by e.g. \citet{2010Roman-Duval} and
\citet{2020Chen}.

\subsection{Aspect ratio and Orientation}
\begin{figure*}
    \begin{center}
    \includegraphics[scale=0.60]{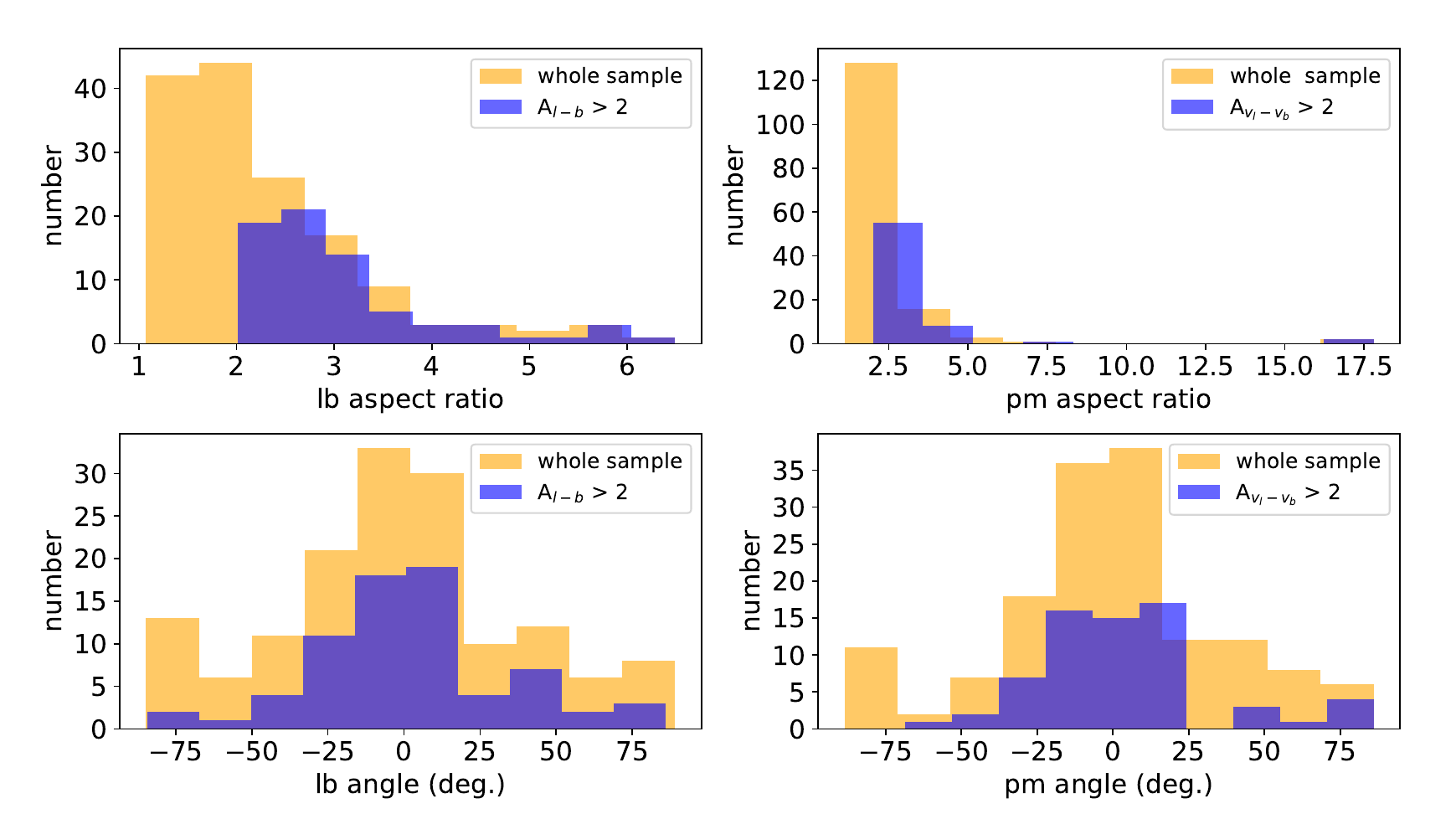}
    \caption{{\bf Distributions of position angles and aspect ratios
    measured in the $l - b$ and $v_l - v_b$ space.} {\bf Upper left:} Distribution of aspect ratio in $l - b$ space. {\bf Lower left:} Distribution of the position angle in $l - b$ space. {\bf Upper right:} Distribution of aspect ratio in $v_l - v_b$ space. {\bf Lower right:} Distribution of the position angle in $v_l - v_b$ space. The Orange histograms
    represent the distribution of the whole 150 structures, the purple
    histograms represent those with aspect ratios greater than 2 in the four
    figures.}
    \label{fig6}
    \end{center}
    \end{figure*}

Our YSO associations are angled mostly parallel to the disk mid-plane. To illustrate this,
we  plot the distributions of the position angle and aspect ratio of the
ellipses respecting the distributions of member YSOs in the $l-b$ and the
$v_l - v_b$ space (Fig.~\ref{fig6}). In the $l-b$ space, the associations appear to be
elongated, with a median aspect ratio of 1.97, and they tend to stay aligned
with the Galactic disk mid-plane, with a median angle of 30$^{\circ}$  with respect to the
 mid-plane. In the $v_l - v_b$ space, the
distribution has a median aspect ratio of 1.82 and a median angle of 27.7$^{\circ}$,
where the velocity dispersion is larger along the Galactic longitude. The difference between the position angle
measured in space and the velocity space is also small, with
a median angle difference of 24.33$^{\circ}$.

\begin{figure*}
    \begin{center}
    \includegraphics[scale=0.61]{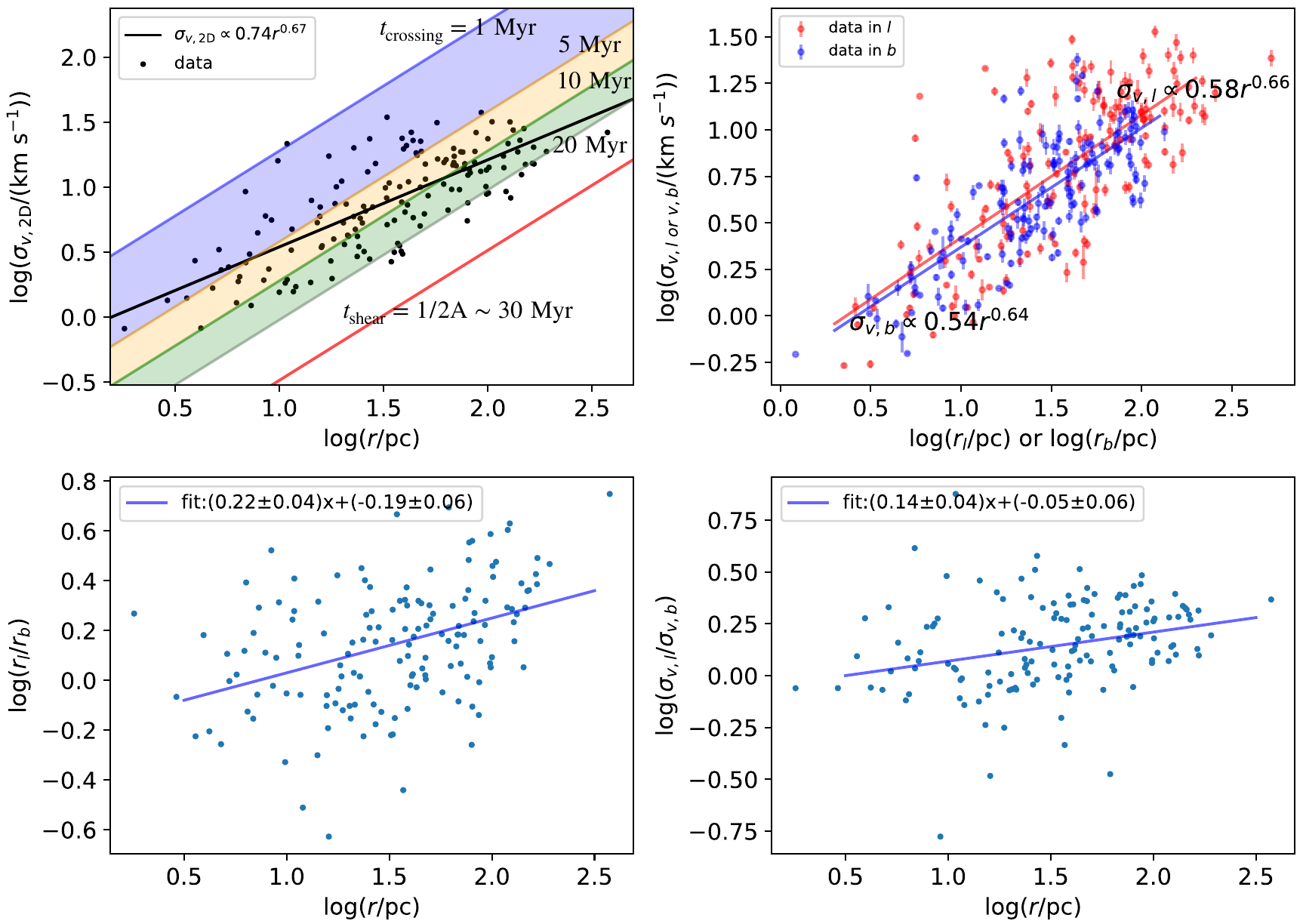}
    \caption{{\bf Relation between velocity and size. Upper left:} ${\rm
    log}(\sigma_{v,{\rm 2D}})$ vs ${\rm log}(r)$. Black dots are the combined 2D
    structure data and the black line is the fitting result to the data points: $y
    = (0.67\pm 0.05)x + (-0.13\pm 0.08)$. The blue, orange, green, and grey lines refer to different crossing times. The red line is the shear time as implied by an Ooct constant A of 16 $\rm km\;s^{-1}\; kpc^{-1}$. {\bf Upper right:} ${\rm
    log}(\sigma_{v_l})$ vs ${\rm log}(r_{l})$ and ${\rm log}(\sigma_{v_b})$ vs
    ${\rm log}(r_{b})$. Red dots are the data in $l$ derection and red line is
    the fitting result: $y = (0.66\pm0.05)x + (-0.24\pm0.08)$. Blue dots are the
    data in $b$ direction and the blue line is the fitting result: $y =
    (0.64\pm0.04)x + (-0.27\pm0.06)$. {\bf Lower left:} ${\rm
    log}(r_l/r_b)$ vs ${\rm log}(r)$. Blue dots are our data and the blue line
    is the fitting result: $y = (0.22\pm0.04)x + (-0.19\pm0.06)$. {\bf Lower
    right:} ${\rm log}(\sigma_{v_l}/\sigma_{v_b})$ vs ${\rm log}(r)$. Blue dots
    are our data and the blue line is the fitting result: $y = (0.14\pm0.04)x +
    (-0.05\pm0.06)$.}
    \label{fig7}
    \end{center}
    \end{figure*}

\subsection{Velocity-Size Relation, and Velocity Anisotropy}
We study the relation between velocity dispersion and size using our YSO
associations. This relation was proposed by \citet{1981Turbulence} and is 
called the Larson relation. In previous studies, the velocity dispersions are  
estimated using the spectral lines from molecules such as the CO, along the
radial direction. We present a first study of the Larson relation using the Gaia
proper motions. This allows us to study the relation
between separation and velocity dispersion measured along $l$ and $b$ directions
separately.

Our result is plotted in Fig.~\ref{fig7}. Along $l$, 
\begin{equation}
    \sigma_{v_l} = 0.58 \;(r_l/{\rm pc})^{0.66}\;({\rm km/s})\,,  
\end{equation}
and  along $b$, 
\begin{equation}
    \sigma_{v_b} =0.54\;(r_b/{\rm pc})^{0.64}\;({\rm km/s}) \;.
\end{equation}
and  combining together using $\sigma_{v\rm,\; 2D} = (\sigma_{v_{\rm min}}^2 +
\sigma_{v_{\rm max}}^2)^{1/2}$, where $\sigma_{v_{\rm max}}$ and $\sigma_{v_{\rm min}}$ are the dispersions measured along the  longer and shorter axes of the velocity dispersion ellipse, thus   
\begin{equation}
\sigma_{v,{\rm 2D}}  = 0.74\;(r/{\rm pc})^{0.67} \;({\rm km/s})\;.
\end{equation}

Our first finding is that the velocity dispersions measured along the $l$ and $b$
directions are almost identical, which means that although the density structure
of the region is anisotropic along $z$ direction and in the $xy$ plane, the turbulence is isotropic. To verify this, we
plot the ratio between the projected length along the Galactic longitude and
latitude direction $r_l / r_b $ again the size $r$,  and as well as the velocity
dispersion ratio $\sigma_{v_l}/\sigma_{v_b}$ against the size $r$ and find the
following relation:

$$r_l/r_b \propto (r/ {\rm pc})^{0.22}\;,$$

Assuming a Larson relation of $\sigma_{\rm v}\propto r^{0.67}$, we have

$$\sigma_{v_l}/\sigma_{v_b} \propto   (r_l/r_b)^{0.22 \times 0.67}  \propto
(r/{\rm pc})^{0.15}\;,$$
 which is consistent with our fitting results: $\sigma_{v_l}/\sigma_{v_b} \propto r^{0.14\pm0.04}$. All the
relations in Fig.~\ref{fig7} are consistent with the turbulence being isotropic.

The scaling exponent 0.67 is steeper than 0.38 in \citet{1981Turbulence} but
close to recent results \citep[e.g.][]{2012MNRAS.425..720S,2013ApJ...779...46H,2017A&A...603A..89K,2018ApJ...860..172S,2021MNRAS.505.4048L}. The slope can be explained if the turbulence is
compressible \citep{2013MNRAS.436.3247K,2021Cen}, or if the turbulence is driven
by accretion and gravitational contraction
\cite[e.g.][]{2011MNRAS.413.2935D,2016ApJ...824...41I,2021Izquierdo}.

We also considered the crossing time for the clouds, which is determined using:
$t_{\rm cross} = 2R/\sigma_{v}$, where R is the radius of the cloud and
$\sigma_v$ is the 2D velocity dispersion. We plotted the lines representing
different crossing times in the upper left panel in Fig.\ref{fig7}. The red
line is the shear time we derived from the Oort constant: $A = 16.31 \pm
0.89$ km s$^{-1}$ kpc$^{-1}$ \citep{2021oort}, which can describe the shearing
caused by the Galactic rotation in the Solar neighborhood
\citep{1927BAN.....3..275O,1986MNRAS.221.1023K}. All our associations have
crossing time much shorter than 30 Myr, implying that the importance of
shear is insignificant compared to that of the turbulence.


\subsection{Energy dissipation rate in different spiral Arms}
In the Kolmogorov theory of turbulence, energy injected into turbulence on large
scales will dissipate on small scales. Following \cite{1999The}, we parameterize
the turbulent energy dissipation rate as  
$\dot{\epsilon}_{\rm kin}  =  \eta {k} \sigma_{v}^3 $, where $\sigma_{v}$ refers to the velocity dispersion of the cloud scale, and ${k}$ is the
driving wavenumber ($k \approx 1/r$). So $ \dot{\epsilon}  =
\sigma_{v}^3/r$ describes the energy dissipation of molecular cloud turbulence, and here we
investigate its variations with respect to their  locations in the Galactic disk.  

Following previous papers \citep{2019MNRAS.487.1400C}, we separate the clouds
into groups associated with different spiral arms (the Sagittarius, Local and
Perseus Arm, see Fig.~\ref{fig5}), and study the energy dissipation rate
measured in terms of $\dot{\epsilon}$.  Since it is hard to resolve small clouds at large distances, for each group, we select median-sized associations ($40 <r <130$ pc), and estimate the energy dissipation rate using $\dot{\epsilon} = \sigma_{v,{\rm 3D}}^3/r$. 
$\sigma_{v,{\rm 3D}}$ is the 3D velocity dispersion estimated using: $\sigma_{v,{\rm 3D}} =
\sqrt{3/2}\sigma_{v,{\rm 2D}}$. Our results are plotted in the lower right panel
of Fig.\ref{fig5}. We note that here we are deriving an un-normalized version of
the energy dissipation rate, as the ``true'' energy dissipation rate is still
dependent on a normalizing factor (\citealt{1995tlan.book.....F}, as summarized by  \citealt{2018qian}), as well as the
efficiency parameter $\eta$, which we can not constrain directly. We find that
the energy dissipation rate decreases with increasing distance from the Galactic
center. By fitting it using an exponential model  we find $\dot{\epsilon} =
1.77\times10^{-4}e^{-0.45 \; (r_{\rm gal} / {\rm kpc})}({\rm erg\ g^{-1}\
s^{-1}})$ where
$r_{\rm gal}$ is the Galactocentric distance, which implies a gradient of  0.20 dex kpc$^{-1}$.

\section{Discussions}

\subsection{Stationary, anisotropic density structures supported isotropic turbulence}
The fact that the associations tend to stay aligned with the disk mid-plane
given an isotropic turbulence despite anisotropic density structures is somewhat
surprising. This is because the very process which leads to the alignment of
associations should also 
leave some imprints on the velocity. One explanation is
that the associations are long-lived entities, with their lifetime as long as the shear time, such that shear can stretch the clouds horizontally, 
aligning the gas associated with the YSOs associated  to the Galactic disk.

The clouds are  thus likely to be objects with $t_{crossing} < t_{\rm
cloud} \approx t_{\rm shear}$, such that shear have sufficient time to align the gas. During the cloud lifetime, turbulence is maintained by a continuous injection. 
In the Solar neighborhood, assuming a Oort
constant of  $A \approx 16 \; \rm km \; s^{-1} \; kpc^{-1}$ \citep{2021oort}, we
expect a cloud lifetime of $t_{\rm cloud} \gtrsim t_{\rm shear}  = \kappa^{-1} =   (1/r| {\rm
d} \Omega/ {\rm d}r| = (2 A)^{-1} = 30\;\rm Myr$ where $\kappa$ is the shear
rate. This is longer than the crossing time of some of the largest association
in our sample, where $t_{\rm cross} \approx 20\;\rm Myr$. This is comparable to
results inferred from the evolution of cloud population evolutions
\citep{2015ApJ...806...72M,2021RNAAS...5..222K}, but is shorter than the
lifetime derived by comparing the locations of $\rm H\,{\sc \rm II}$ regions
with the locations of clouds   \citep{2019Kruijssen}.

These results also provide insights into the nature of turbulence in clouds.
There are different schools of thought concerning the link between density
and velocity structures. The first believes that turbulence creates the density
structure upon which collapse occurs \citep{2017Semadeni}. \cite{2002Padoan}
found the slope of the stellar (Initial Mass Function) IMF can be predicted and
produced by the turbulence. The second believes that the role of turbulence is
to provide support, where only gravitationally bound density fluctuations can
collapse \citep{2008ApJ...684..395H,2013MNRAS.430.1653H}. The third,
particularly \cite{2003Maclow} believes that the role of supersonic turbulence
is to provide support, leading to slower collapse and longer cloud lifetime. Our
results favor the third one. The density structures appear to be supported by
turbulence and long-lived. 


\subsection{Energy source of molecular clouds}
Our results indicate that the energy dissipation rate of turbulence decreases
with the Galactocentric distance. This can be explained if the turbulence is
driven by molecular cloud coagulations. Along this line of thought,
\cite{2010Klessen} and \cite{2010Elmegreen} argued
that accretion can explain the observed level of turbulence at the Galactic,
molecular cloud and core scales. Using an analytical model and numerical
simulations, \cite{2011MNRAS.413.2935D}, \cite{2015Goldbaum} and \cite{2016Juan} demonstrated that a large cloud can
maintain its turbulence by accreting smaller clouds.

Our conclusion results from the fact that molecular clouds are more turbulent in gas-rich  regions in the Milky Way disk as characterized by $\Sigma_{\rm disk}$. This fact can be
explained in the collision-driven turbulence scenario, where regions of higher densities have higher cloud-cloud collision rates, leading to more turbulence. Quantitatively, 
\cite{2017Li} studied the co-evolution between the Galactic disk and molecular cloud, and proposed the following equation linking the properties of the disk to the energy dissipation rate of the molecular cloud:
\begin{equation}\label{eq:li}
 \sigma_{v,{\rm cloud}} \approx \sigma_{v,{\rm disc}}(\frac{\Sigma_{{\rm disc}}}{\Sigma_{\rm cloud}})^{1/3}(\frac{r_{\rm cloud}}{H_{{\rm disc}}})^{1/3}
\end{equation}
where $\Sigma_{\rm cloud} \approx 10 M_{\odot}\;\rm pc^{-2}$ and is the mean
surface density of the cloud, $H_{{\rm disc}}$ is the scalehight of the disc.
The level of turbulence measured in terms of $\dot{\epsilon} \propto
\sigma_{v,{\rm 3D}}^3/r$ is proportional to $\Sigma_{{\rm disc}} / \Sigma_{\rm
cloud}$, as an increased disk surface density leads to higher chances of cloud
collisions. We note for this calculation to be valid, the energy injection rate
should follow $\dot \epsilon  = \sigma_{\rm v, disk}^2 t_{\rm collisions}$.
This, we believe, is a reasonable estimate of the energy injection rate, and and
the results should hold  whether the clouds are long-live or not.
This is consistent with conclusions from simulation  of  barred and spiral galaxies
where clouds can be relatively short lived and still collide
\citep{2014MNRAS.439..936F,2015MNRAS.446.3608D}.  We also note that results is
inconsistent with the scenario of molecular cloud turbulence driven directly by stellar feedback.  This is because in that case the energy injection rate should be related to $\Sigma_{SF} / \Sigma_{\rm gas}$, where
$\Sigma_{\rm SF} $ is the surface density of star formation tracers, and
$\Sigma_{\rm gas}$ is the surface density of gas. To explain the varying 
turbulence as a function of location, we expect the star formation efficiency to
vary radially, which is inconsistent with observational results which point to a
constant gas depletion time \citep{2013AJ....146...19L}.

\cite{Marc2016PHYSICAL} measured the surface density distribution of the Milky Way disk as a function of Galactocentric radius, and obtained the following equation:
\begin{equation}
\Sigma_{{\rm disk}} \propto e^{-0.5\; r_{\rm gal}} \;.
\end{equation}
Neglecting the radial variations of $H_{\rm disk}$ and $\sigma_{\rm v, disk}$,
according to Eq. \ref{eq:li}, we expect 
\begin{equation}
 \dot \epsilon = \eta \sigma_{\rm v}^3 / l \propto e^{-0.5\; r_{\rm gal}}\;, 
\end{equation}
which agrees with our result, where $\dot{\epsilon} \propto e^{-0.45\; r_{\rm
gal}}$. The decreasing energy dissipation rate of the molecular cloud
turbulence with increasing $r_{\rm gal}$ is likely caused by the decrease
of surface density of the molecular disc, which leads to a lower collisional rate.
We note that in our picture, we focus on the change of the cloud-averaged energy dissipation rate. Small(sub-kpc)-scale  fluctuations of surface density might lead to local variations in the energy dissipation, which is averaged out after we have divided our clouds into groups.

\section{Conclusions}
Using a sample of 15149 YSOs younger than 3 Myr, we extract 150 associations with
sizes ranging from several pc to one hundred pc and study their velocity
structures. This allows us to probe the velocities of the gas that these YSOs are associated with.
 Our major results include:
\begin{enumerate}
 \item All the structures are elongated and are orientated parallel to the
 Galactic mid-plane. 
 \item We find that the velocity  dispersions and sizes are related by $\sigma_{v,{\rm 2D}} = 0.74 {\;}(r/{\rm
 pc})^{0.67\pm0.05}({\rm km\ s^{-1}})$, which is often called the Larson relation. The exponent 0.67 is consistent with
 recent simulation results.
 
 \item Using proper motion measurements, we study the velocity dispersion-size
 relation measured along the Galactic longitude and latitude separately. Along
 $l$ direction, we obtain $\sigma_{v_l} = 0.58 {\;} (r_l/{\rm pc})^{0.66}({\rm
 km\ s^{-1}})$ and along $b$ direction, we obtain $\sigma_{v_b} = 0.54 {\;}
 (r_b/{\rm pc})^{0.64}({\rm km\ s^{-1}})$. The results indicate that turbulence
 in molecular clouds is isotropic.
 \item Although the density structure is anisotropic, the turbulence is
 consistent with being isotropic. This can be explained if molecular clouds are
 long-lived ($t_{\rm cloud} \gtrsim t_{\rm shear}$), such that their overall density structures are shaped by Galactic-scale
 process such as shear. During the cloud lifetime, turbulence is maintained by a continuous injection. 

 \item We compute the energy dissipation rate of turbulence using
 $\sigma_{v,{\rm 3D}}^3/r$, and find it depends on the distance from the
 Galactic center $r_{\rm gal}$, where $\dot{\epsilon} =
 1.77\times10^{-4}e^{-0.45(r_{\rm gal}/{\rm (pc)})}({\rm erg\ g^{-1}\ s^{-1}})$, where
 a slope of 0.20 dex kpc$^{-1}$ is implied.

 \item The decreasing energy dissipation rate with increasing Galactocentric
 radius can be explained by a reduction of clouds collisions rates, as the clouds located in the inner Galaxy where the 
 disk surface densities are higher have higher changes of accreting smaller clouds. 
\end{enumerate}

\section*{Acknowledgements}We thank our referees whose comments helped to improve our paper significantly.
This work is partially supported by the Postgraduate's Research and Innovation Project of Yunnan University (No. 2019236). GXL acknowledges supports from 
NSFC grant W820301904 and 12033005. BQC is supported by the National Key R\&D Program of China No. 2019YFA0405500, National Natural Science Foundation of China 12173034, 11803029 and 11833006, and the science research
grants from the China Manned Space Project with NO.\,CMS-CSST-2021-A09, CMS-CSST-2021-A08 and CMS-CSST-2021-B03.

This work presents results from the European Space Agency (ESA) space mission Gaia. Gaia data are being processed by the Gaia Data Processing and Analysis Consortium (DPAC). Funding for the DPAC is provided by national institutions, in particular the institutions participating in the Gaia MultiLateral Agreement (MLA). The Gaia mission website is https://www.cosmos.esa.int/gaia. The Gaia archive website is https://archives.esac.esa.int/gaia. This research has used ASTRODENDRO, a
PYTHON package to compute dendrograms of Astronomical data
(http://www.dendrograms.org/).

\section*{DATA AVAILABILITY}
The paper makes use of published data from \citet{2016An} and Gaia DR2 \citep{GAIADR2}. Our Table \ref{tab1} will be made available online upon publication.
\bibliographystyle{mnras}
\bibliography{YSO.bib}

\appendix
\section{Parameters of Associations}
In Table.\ref{tab1}, we list the properties of the assocations, including the
mean longitude, the mean latitude, the number of the member stars, the mean
distance, the position angle, and aspect ratio in $l - b$ and $v_l-v_b$ space,
the angle difference between $l - b$ and $v_l-v_b$ space, the velocity
dispersion in $l$ and $b$ directions, the combined 2D velocity dispersion, the
size in $l$ and $b$ directions, and the combined size of some of the 150
associations. 

\clearpage
\onecolumn

\begin{landscape}
\begin{longtable}{rrrrrrrrrrrrrrrr}
\caption{Parameters of YSO associations (only a fraction of the table is shown and the full table can be downloaded online\url{http://paperdata.china-vo.org/zjx/yso_association/upload_association_para.csv }).}
\label{tab1}\\

\hline
id  & $l(^{\circ})$ & $b(^{\circ})$ & ${\rm n}_{\rm star}$ & d (kpc) & PA$_{v_l-v_b}(^{\circ})$ & A$_{v_l-v_b}$ & PA$_{l-b}(^{\circ})$ & A$_{l-b}$ & $\Delta_{\rm A}$($^{\circ}$) & $\sigma_{v_l}$(km/s) & $\sigma_{v_b}$(km/s)  & $\sigma_{v,{\rm 2D}}$(km/s) & $r_l$(pc) & $r_b$(pc)  & $r$(pc) \\ \hline
\endfirsthead
\endhead
\hline
\endfoot
\endlastfoot
G171.700-015.500 & 171.67  & -15.461 & 62  & 0.13 & -41.33 & 1.7  & 10.25  & 2.89 & 51.57 & 1.43  & 1.33  & 1.61  & 14.36  & 5.6    & 10.9   \\
G169.600-015.700 & 169.56  & -15.704 & 33  & 0.13 & -33.19 & 1.62 & 11.71  & 2.87 & 44.9  & 1.46  & 1.21  & 1.6   & 8.32   & 3.37   & 6.35   \\
G168.700-015.900 & 168.681 & -15.883 & 24  & 0.13 & -64.2  & 1.3  & -84.77 & 1.17 & 20.57 & 0.9   & 1.03  & 0.99  & 2.67   & 3.12   & 2.9    \\
G174.500-015.000 & 174.466 & -14.997 & 45  & 0.14 & 57.49  & 1.79 & -63.54 & 1.66 & 58.96 & 1.31  & 1.61  & 1.13  & 5.47   & 7.3    & 6.45   \\
G338.205+009.000 & 338.205 & 8.989   & 47  & 0.16 & -25.54 & 2.96 & 14.48  & 6.49 & 40.03 & 1.67  & 0.96  & 0.92  & 11.4   & 3.43   & 8.42   \\
G336.422+008.600 & 336.422 & 8.562   & 21  & 0.16 & -25.4  & 4.07 & 28.86  & 2.59 & 54.26 & 2.41  & 1.27  & 1.23  & 4.65   & 3.06   & 3.94   \\
G339.665+009.300 & 339.665 & 9.346   & 24  & 0.16 & -71.61 & 1.22 & 3.45   & 1.87 & 75.06 & 0.54  & 0.62  & 0.53  & 2.25   & 1.21   & 1.81   \\
G352.880+019.500 & 352.88  & 19.484  & 255 & 0.14 & -76.84 & 1.43 & -84.38 & 2.04 & 7.54  & 1.12  & 1.5   & 1.13  & 8.94   & 17.87  & 14.13  \\
G353.255+018.900 & 353.255 & 18.906  & 166 & 0.14 & -76.76 & 1.49 & 86.0   & 3.33 & 17.24 & 1.01  & 1.41  & 1.03  & 5.0    & 16.21  & 11.99  \\
G353.199+017.000 & 353.199 & 17.012  & 107 & 0.14 & -76.26 & 1.23 & 82.05  & 1.86 & 21.68 & 1.1   & 1.29  & 1.14  & 3.28   & 5.92   & 4.79   \\
G353.383+022.500 & 353.383 & 22.473  & 68  & 0.14 & -84.42 & 1.36 & 24.8   & 1.55 & 70.78 & 0.79  & 1.04  & 0.81  & 7.0    & 5.33   & 6.23   \\
G297.100-015.200 & 297.064 & -15.248 & 59  & 0.19 & -15.3  & 1.27 & -71.88 & 1.97 & 56.58 & 1.12  & 0.91  & 1.07  & 2.6    & 4.37   & 3.6    \\
G303.600-014.700 & 303.583 & -14.666 & 24  & 0.2  & 54.32  & 1.71 & 70.31  & 1.91 & 15.98 & 0.55  & 0.63  & 0.64  & 3.14   & 5.04   & 4.2    \\
G158.400-020.700 & 158.446 & -20.742 & 49  & 0.3  & -12.48 & 1.47 & 41.79  & 1.67 & 54.27 & 2.05  & 1.42  & 2.03  & 5.86   & 5.55   & 5.71   \\
G160.300-017.900 & 160.342 & -17.938 & 33  & 0.32 & 12.45  & 2.42 & 24.01  & 1.46 & 11.56 & 3.02  & 1.41  & 2.37  & 5.71   & 4.47   & 5.13   \\
G104.424+014.000 & 104.424 & 14.017  & 18  & 0.33 & 10.7   & 1.36 & -21.06 & 2.92 & 31.76 & 1.0   & 0.77  & 0.81  & 9.22   & 4.71   & 7.32   \\
G208.400-018.400 & 208.356 & -18.439 & 525 & 0.39 & 43.87  & 1.11 & -34.06 & 1.98 & 77.94 & 2.18  & 2.18  & 1.96  & 43.11  & 34.42  & 39.01  \\
G210.100-019.500 & 210.082 & -19.484 & 285 & 0.39 & -8.2   & 1.31 & 0.16   & 2.64 & 8.36  & 2.15  & 1.66  & 1.77  & 23.35  & 8.84   & 17.66  \\
G206.000-016.300 & 205.964 & -16.34  & 155 & 0.4  & 60.32  & 1.36 & -68.98 & 1.64 & 50.7  & 2.38  & 2.77  & 2.28  & 16.68  & 23.74  & 20.52  \\
G206.400-017.000 & 206.432 & -16.955 & 79  & 0.4  & 61.76  & 1.62 & 13.14  & 1.29 & 48.62 & 2.01  & 2.59  & 2.09  & 12.62  & 10.05  & 11.41  \\
...         & ...           & ...             & ...              & ...             & ...                      & ...                      & ...              & ...                       & ...                & ...         & ...         & ...             & ...         & ...         & ...             \\
G345.218+000.800 & 345.218 & 0.778  & 89      & 1.59 & -36.42   & 1.82     & -19.67 & 2.71   & 16.75     & 13.29     & 11.31     & 14.17      & 149.76 & 76.25 & 118.83 \\
G344.161+001.200 & 344.161 & 1.222   & 60  & 1.57 & -41.15 & 2.44 & -31.19 & 1.34 & 9.96  & 12.12 & 10.94 & 13.4  & 67.09  & 58.77  & 63.07  \\
G352.350+002.800 & 352.35  & 2.8     & 42  & 1.07 & -14.24 & 3.12 & 15.38  & 1.66 & 29.61 & 18.38 & 7.47  & 15.2  & 81.2   & 53.16  & 68.63  \\
G335.700-001.500 & 335.746 & -1.516  & 29  & 1.18 & -18.6  & 1.57 & -13.47 & 4.44 & 5.12  & 11.4  & 8.05  & 9.85  & 103.57 & 34.02  & 77.09  \\
G333.948+000.300 & 333.948 & 0.304   & 45  & 2.14 & 9.22   & 2.69 & -11.5  & 1.81 & 20.72 & 29.47 & 11.89 & 29.64 & 156.06 & 91.16  & 127.8  \\
G336.700-001.800 & 336.745 & -1.756  & 24  & 1.17 & 51.3   & 1.58 & 8.57   & 1.97 & 42.73 & 4.41  & 4.87  & 5.52  & 32.45  & 17.13  & 25.95  \\
G333.892+000.200 & 333.892 & 0.196   & 33  & 2.22 & 6.82   & 1.96 & -10.0  & 1.94 & 16.83 & 24.49 & 12.75 & 24.12 & 164.54 & 89.27  & 132.37 \\
G304.110+000.600 & 304.11  & 0.554   & 80  & 2.12 & -2.5   & 1.96 & -12.91 & 2.67 & 10.4  & 25.21 & 12.79 & 25.12 & 193.07 & 84.52  & 149.03 \\
G307.000-000.000 & 307.05  & -0.011  & 15  & 2.17 & -74.32 & 1.21 & -14.62 & 1.65 & 59.7  & 20.44 & 23.93 & 22.11 & 67.57  & 44.08  & 57.04  \\
G303.056+000.800 & 303.056 & 0.804   & 45  & 1.97 & 1.84   & 3.09 & 3.25   & 1.44 & 1.42  & 24.95 & 8.18  & 24.99 & 102.03 & 70.97  & 87.88  \\
G303.191+001.100 & 303.191 & 1.128   & 29  & 1.83 & 16.21  & 1.69 & 1.25   & 3.59 & 14.96 & 12.92 & 8.24  & 12.95 & 105.7  & 29.52  & 77.6   \\
G302.240+001.100 & 302.24  & 1.139   & 18  & 1.82 & 15.23  & 1.84 & 12.35  & 2.27 & 2.88  & 16.0  & 9.5   & 15.72 & 57.55  & 28.22  & 45.32  \\
G294.900-001.400 & 294.916 & -1.392  & 34  & 2.42 & 13.43  & 1.81 & -78.52 & 1.24 & 88.05 & 13.61 & 7.96  & 13.38 & 61.76  & 75.06  & 68.73  \\
G290.577+000.400 & 290.577 & 0.422   & 62  & 2.44 & -4.39  & 1.53 & -77.95 & 1.43 & 73.55 & 13.2  & 8.53  & 13.22 & 71.74  & 98.9   & 86.39  \\
G290.600-000.100 & 290.555 & -0.118  & 39  & 2.5  & -3.26  & 1.47 & -17.88 & 1.71 & 14.62 & 12.43 & 8.33  & 12.59 & 80.48  & 52.72  & 68.03  \\
G283.200-001.900 & 283.201 & -1.877  & 38  & 2.53 & 14.37  & 3.78 & 51.64  & 1.08 & 37.27 & 18.01 & 6.52  & 18.14 & 85.97  & 87.49  & 86.74  \\
G286.900-000.700 & 286.906 & -0.666  & 27  & 2.63 & -20.02 & 1.78 & -13.04 & 2.43 & 6.98  & 15.91 & 10.33 & 14.88 & 95.0   & 44.67  & 74.23  \\
G086.975+003.900 & 86.975  & 3.932   & 16  & 1.62 & 18.88  & 1.7  & 47.99  & 2.06 & 29.11 & 22.07 & 14.7  & 18.08 & 41.54  & 44.31  & 42.94  \\
G068.600-001.200 & 68.587  & -1.207  & 18  & 1.87 & 7.89   & 1.14 & 2.81   & 3.94 & 5.08  & 10.3  & 8.95  & 10.29 & 134.89 & 34.88  & 98.52              \\ \hline

\end{longtable}
\end{landscape}

\clearpage
\twocolumn

\end{document}